\documentstyle[epsfig,prl,twocolumn,aps]{revtex} 
\begin{document}
%
%
\newcommand{\be}[1]{\begin{equation} \label{(#1)}}
\newcommand{\ee}{\end{equation}}
\newcommand{\ba}[1]{\begin{eqnarray} \label{(#1)}}
\newcommand{\ea}{\end{eqnarray}}
\newcommand{\nn}{\nonumber}
\newcommand{\rf}[1]{(\ref{(#1)})}
%
%
\def\nm{\hbox{$\nu_\mu$ }}
\def\nt{\hbox{$\nu_\tau$ }}
\def\rp{$R_p \hspace{-1em}/\;\:$ }
\def\lv{$L \hspace{-0.6em}/\;\:$ }
\def\rpm{R_p \hspace{-0.8em}/\;\:}
\def\rpt{$R_p \hspace{-0.85em}/\ \ $}
\def \znbb {\beta\beta_{0\nu}}
\def\rve{\langle {\tilde \nu_e}\rangle}
\def\rvm{\langle {\tilde \nu_{\mu}}\rangle}
\def\rvt{\langle {\tilde \nu_{\tau}}\rangle}
\def\thefootnote{\fnsymbol{footnote}}
\newcommand{\ptmis}{{ {\rm p} \hspace{-0.53 em} \raisebox{-0.27 ex} {/}_T }}
 \def\mathrm#1{\mbox{\rm #1}}
\def\bold#1{\setbox0=\hbox{$#1$}%
     \kern-.025em\copy0\kern-\wd0
     \kern.05em\copy0\kern-\wd0
     \kern-.025em\raise.0433em\box0 }
\def\21{$SU(2) \otimes U(1)$}
\newcommand{\matriz}{\left[\begin{array}} 
\newcommand{\finmatriz}{\end{array}\right]} 
\def\frad#1#2{\frac{\displaystyle{#1}}{\displaystyle{#2}}}
\def\frap#1#2{{\hbox{$\frac{#1}{#2}$}}}
\def\half{{\textstyle{1 \over 2}}}
\def\eighth{{\textstyle{1 \over 8}}}
\def\etal{\hbox{\it et al., }}
\def\eq#1{{eq. (\ref{#1})}}
\def\sm{\hbox{Standard Model }}
\def\VEV#1{\left\langle #1\right\rangle}
\def\lsim{\raise0.3ex\hbox{$\;<$\kern-0.75em\raise-1.1ex\hbox{$\sim\;$}}}
\def\gsim{\raise0.3ex\hbox{$\;>$\kern-0.75em\raise-1.1ex\hbox{$\sim\;$}}}
\def\mpl#1#2#3{          {\it Mod. Phys. Lett. }{\bf #1} (19#2) #3}
\def\np#1#2#3{           {\it Nucl. Phys. }{\bf #1} (19#2) #3}
\def\nps#1#2#3{          {\it Nucl. Phys. B (Proc. Suppl.) }
    {\bf #1} (19#2) #3}
\def\pl#1#2#3{           {\it Phys. Lett. }{\bf #1} (19#2) #3}
\def\ppnp#1#2#3{           {\it Prog. Part. Nucl. Phys. }{\bf #1} (19#2) #3}
\def\pr#1#2#3{           {\it Phys. Rev. }{\bf #1} (19#2) #3}
\def\prep#1#2#3{         {\it Phys. Rep. }{\bf #1} (19#2) #3}
\def\prl#1#2#3{          {\it Phys. Rev. Lett. }{\bf #1} (19#2) #3}
\def\zp#1#2#3{          {\it Z. Phys. }{\bf #1} (19#2) #3}
\twocolumn[\hsize\textwidth\columnwidth\hsize\csname @twocolumnfalse\endcsname
\title{A Supersymmetric Solution to the Solar and Atmospheric Neutrino 
Problems}
\author{J.~C. Rom\~ao~${}^1$, M.A. D\'{\i}az~${}^2$, M. Hirsch~${}^3$,  
W. Porod~${}^3$, and J.~W.~F Valle~${}^3$}
\address{${}^1$Departamento de F\'\i sica, Instituto Superior T\'ecnico\\ 
             A. Rovisco Pais, 1049-001 Lisboa, Portugal\\
${}^2$High Energy Physics, Florida State University,
Tallahassee, Florida 32306, USA\\
${}^3$Instituto de F\'{\i}sica Corpuscular - IFIC/CSIC,
             Departamento de F\'{\i}sica Te\`orica \\
             Universitat de Val\`encia, 46100 Burjassot, 
             Val\`encia, Spain}
\maketitle
\begin{abstract} 
The simplest unified extension of the Minimal Supersymmetric Standard
Model with bi-linear R--Parity violation provides a predictive scheme
for neutrino masses which can account for the observed atmospheric and
solar neutrino anomalies in terms of bi-maximal neutrino mixing. The
maximality of the atmospheric mixing angle arises dynamically, by
minimizing the scalar potential, while the solar neutrino problem can
be accounted for either by large or by small mixing oscillations.  One
neutrino picks up mass by mixing with neutralinos, while the
degeneracy and masslessness of the other two is lifted only by loop 
corrections.
Despite the smallness of neutrino masses R-parity violation is
observable at present and future high-energy colliders, providing an
unambiguous cross-check of the model.
\end{abstract}
\vskip0pc]



The pattern of fermion masses and mixings constitutes one of the most
important issues in modern physics. Here we propose a model for the
structure of lepton mixing which accounts for the atmospheric and
solar neutrino anomalies. It is based on the simplest one-parameter
extension of minimal supergravity with bi-linear R--Parity violation
\cite{epsrad} as would arise, perhaps, from gravitation. 

The recent announcement of high statistics atmospheric neutrino data
by the SuperKamiokande collaboration \cite{Fukuda:1998mi} has
confirmed the deficit of muon neutrinos, especially at small zenith
angles, opening a new era in neutrino physics. 
Although there may be alternative solutions of the atmospheric
neutrino anomaly ~\cite{Gonzalez-Garcia:1998hj} it is fair to say that
the simplest interpretation of the data is in terms of \nm to \nt
flavour oscillations with maximal mixing. This excludes a large mixing
among $\nu_{\tau}$ and $\nu_e$~\cite{Fukuda:1998mi}, in agreement also
with the Chooz reactor data.   On the other hand the
persistent disagreement between solar neutrino data and theoretical
expectations~\cite{BP98} has been a long-standing problem in physics. 
Recent solar neutrino
data~\cite{Smy:1999tt} are consistent with both
vacuum oscillations and MSW conversions. In the latter case one can
have either the large or the small mixing angle solutions, with a
slight trend towards the latter~\cite{MSW99}. The situation might
become clearer in the near future when rate-independent observables
such as spectrum, day-night and seasonal variations are better
measured. In summary one sees that while quarks are weakly mixed,
there is now the intriguing possibility that neutrino mixing is (close
to) bi-maximal~\cite{bimax}.

Our model breaks lepton number and therefore necessarily generates
non-zero Majorana neutrino masses~\cite{Schechter:1980gr}. It has
strong predictive power and allows for a dynamical determination of
the atmospheric neutrino angle. Moreover it leads, under certain
circumstances, to bi-maximal neutrino mixing. At tree-level only one
of the neutrinos picks up a mass by mixing with
neutralinos~\cite{Ross:1985yg}, leaving the other two neutrinos
massless ~\cite{ProjectiveMassMatrix}. While this can explain the
atmospheric neutrino problem, to reconcile it with the solar neutrino
data requires going beyond the tree-level approximation. This is the
purpose of the present paper. For an analysis including only the
atmospheric neutrino problem in the tree-level approximation see
ref.~\cite{Bednyakov:1998cx}.

We have performed a full one-loop calculation of the
neutralino-neutrino mass matrix in the bi-linear \rp MSSM. As is shown
below, in order to explain the solar and atmospheric neutrino data it
is both necessary and sufficient to work at one-loop level.  In
contrast to other papers \cite{Hempfling,allothers} we have taken
special care to achieve gauge invariance of the calculation. Moreover
we have performed the renormalization of the heaviest neutrino, thus
refining the approximate approaches used, for example, in
ref.~\cite{allothers}. For estimates in the approximation where loop
neutrino masses arise just from tri-linear R--parity breaking see ref.
\cite{allothers3}. 


Bilinear R-parity breaking supersymmetry has been extensively
discussed in the literature~\cite{epsrad}. It is motivated on the one
hand by the fact that it provides an effective truncation of models
where R--parity breaks spontaneously \cite{Masiero:1990uj} around the
weak scale. Moreover, they allow for the radiative breaking of
R-parity, opening also new ways to unify Gauge and Yukawa
couplings~\cite{Diaz:1998wz} and with a potentially slightly lower
prediction for $\alpha_s$~\cite{Diaz:1999is}. If present at the
fundamental level tri--linear breaking of R--parity will always imply
bi-linear breaking at some level, as a result of the renormalization
group evolution. In contrast, bi-linear breaking may exist in the
absence of tri--linear, as would be the case if it arises
spontaneously.

Here, we concentrate only on those features of the model which
are related to neutrino masses.  Our model consists of the MSSM
particle spectrum and superpotential except for the addition of the
following $\epsilon_i$ terms

\be{SuperPot}
W = W_{MSSM} + \epsilon_i\widehat L_i^a\widehat H_u^b .
\ee
Should supersymmetry not be broken, the above bi-linear terms would be
superfluous since a suitable redefinition of the lepton and Higgs
superfields \cite{HS84} would convert them into trilinear R-parity
violating terms. However, since supersymmetry must be broken, they
give rise to a second source for R-parity violation:
\be{SoftBreaking}
V_{soft} = V_{soft,MSSM} +B_i\epsilon_i\widetilde L_i^a H_u^b
\ee
In the presence of soft supersymmetry breaking terms, the bi-linear
R-parity violating terms can not be rotated away, except in the
particular case when $B=B_i$ and $m_{H_d}^2=m_{L_i}^2$, i=1,2,3 which
is untypical. Thus we prefer to work in the original basis, containing
no trilinear \rp vertices.

The presence of the bi-linear terms in \rf{SoftBreaking} imply that
the tadpole equations for the sneutrinos are non-trivial, i.e. lead to
finite VEV for the scalar neutrinos. As a consequence the neutrinos
and neutralinos, charged leptons and charginos as well as the Higgses
and sleptons of the MSSM mix with each other.  Detailed mass matrices
are found in \cite{BRPVMassMatrices,Paper2}. For the neutrino masses
the most important aspect is, of course, the neutrino-neutralino
mixing. It generates the following ($7\times 7$) mass matrix,

\be{nmm}
{\cal M}_0 =  \left(
                    \begin{array}{cc}
                    0 & m \\
                    m^T & {\cal M}_{\chi^0} \\
                    \end{array}
              \right),
\ee
where ${\cal M}_{\chi^0}$ is the usual MSSM neutralino mass matrix 
and the sub-matrix $m$ contains entries from the bi-linear \rp 
parameters,

\be{bnmm}
m =   \left(
            \begin{array}{cccc}
     -\frac{1}{2}g'\rve & \frac{1}{2}g\rve & 0 & \epsilon_e \\
     -\frac{1}{2}g'\rvm & \frac{1}{2}g\rvm & 0 & \epsilon_{\mu} \\
     -\frac{1}{2}g'\rvt & \frac{1}{2}g\rvt & 0 & \epsilon_{\tau} \\
                    \end{array}
              \right),
\ee
and $\rve$, $\rvm$ and $\rvt$ are the VEVS of the scalar neutrinos
and $g,\:g'$ are electroweak gauge couplings.

It is easy to show that this mass matrix \rf{nmm} has such a structure
that only one combination of $\nu_e$, $\nu_{\mu}$, $\nu_{\tau}$ picks
up a mass, while the remaining two states remain massless.  This
structure is reminiscent of that found in
ref.\cite{ProjectiveMassMatrix}. If the RPV parameters are smaller
than the typical size of the MSSM parameters, there exists a simple
approximation formula for the non-zero mass of the neutrino,

\be{nonzero}
m_{\nu} \simeq  
\frac{M_1 g^2 + M_2 {g'}^2}{4 det({\cal M}_{\chi^0})} 
|{\vec \Lambda}|^2,
\ee
where, 
\be{deflam}
\Lambda_i = \mu \langle {\tilde \nu}_i\rangle + v_d \epsilon_i, 
\ee
and $M_1,\:M_2$ are supersymmetry breaking electroweak gaugino masses.
This ``alignment'' vector plays a prominent role in all the discussion
below since it will fix both the overall neutrino mass scale as well
as the atmospheric neutrino mixing. With two neutrinos being massless
one of the angles describing the mixing between them can be rotated
away~\cite{Schechter:1980bn}. However, in the presence of loops this
angle, which will characterize the solar neutrino conversions, will
acquire a meaning, together with a (Dirac-type) CP phase.

There are three simple topologies of relevant Feynman diagrams
contributing to the neutrino-neutralino mass matrix~\cite{Paper2}.
With these the one-loop corrected mass matrix is calculated as,
\ba{1loopMass}\nn
M_{ij}^{pole} & =& M_{ij}^{\overline {DR}}(\mu_R) 
              + \frac{1}{2} \Big(
                \Pi_{ij}(p_i^2) + \Pi_{ij}(p_j^2) \\
              &  - & m_{\chi^0_i} \Sigma_{ij}(p_i^2) - 
                  m_{\chi^0_j} \Sigma_{ij}(p_j^2) \Big)
\ea
where $\Sigma_{ij}$ and $\Pi_{ij}$ are self-energies.  For a complete
description see~\cite{Paper2}. Here, ${\overline {DR}}$ signifies the
minimal dimensional reduction subtraction scheme and $\mu_R$ is the
renormalization scale.  In order to check for gauge invariance in
calculating $\Pi_{ij}$ and $\Sigma_{ij}$ we have used the general
$R_{\xi}$ gauges. As demonstrated in ref.~\cite{Paper2} gauge
invariance requires the inclusion of the tadpole diagrams for the
Goldstone bosons associated with the $Z^0$ and $W^\pm$ into the self
energies. Moreover, in minimizing the scalar potential, for
consistency reasons it is necessary to also include tadpole diagrams,
when solving the tadpole equations, but excluding the Goldstone
tadpole graphs which have been already included into the self energies
\cite{Paper2}. On the other hand, if Goldstone tadpole graphs are kept
in the tadpole equations rather than in the self energies, gauge
dependent VEVs would be generated. This problem has been ignored so
far in all previous descriptions~\cite{allothers}.

The scalar potential contains terms linear in the real part of the
neutral scalar fields $\sigma_{\alpha}\equiv 
(\sigma^0_d,\sigma^0_u, Re(\tilde\nu_1),Re(\tilde\nu_2),Re(\tilde\nu_3))$
\begin{equation}
V_{linear}=t_d\sigma^0_d+t_u\sigma^0_u+t_i Re(\tilde\nu_{i})
\equiv t_{\alpha}\sigma^0_{\alpha}
\end{equation}
The coefficients of these terms are the tadpoles.  Including the
one--loop contribution we write
\begin{eqnarray}
t_{\alpha}&=&t^0_{\alpha} -\delta t^{\overline{DR}}_{\alpha}
+T_{\alpha}(Q)\\ \nonumber
&=&t^0_{\alpha} +T^{\overline{DR}} _{\alpha}(Q)
\label{tadpoles}
\end{eqnarray}
where $T^{\overline{DR}} _{\alpha}(Q)\equiv -\delta
t^{\overline{DR}}_{\alpha} +T_{\alpha}(Q)$ are the finite one--loop
tadpoles. The minimization of the scalar potential corresponds then to
solve $t_{\alpha}=0$.  This is done by solving these equations for the
soft masses squared. This is easy because those equations are linear
on the soft masses squared. However the values obtained in this way,
which we call $m^2_i$, are not equal to the values $m^2_i(RGE)$ that
we got via the Renormalization Group Equations (RGE) starting from
universal soft masses at the unification scale.  To achieve equality
we define a function
\begin{equation}
\eta= max \left(  \frac{m^2_i}{m^2_i(RGE)},
\frac{m^2_i(RGE)}{m^2_i}
\right) \quad \forall i 
\end{equation}
with the obvious property that $\eta \ge 1$.
Then we adjust the parameters at unification scale to minimize $\eta$.


We have performed a complete scan of the \rp MSSM parameter space,
following the procedure outlined above. As an example we allow the MSSM
parameters to vary within the range $M_2, |\mu|$ up to 500 GeV, $m_0$
up to 1 TeV, and assumed $|A_0/m_0| \le 3$, which helps avoiding
charge breaking minima. Moreover we assumed $\tan\beta \lsim 10$. The
latter is needed in order to obtain a nearly maximal atmospheric
angle, since otherwise the sizeable loop involving down quarks and
squarks would distort this feature due to its very strong $\tan\beta$
dependence.
Note also that the bound on $\tan\beta$ implies that the lightest CP
even Higgs boson mass lies below 115 GeV or so.  As we will see below
we can find simultaneous solutions to the atmospheric and solar
neutrino problems only in those parts of parameter space where the
one-loop contributions to the {\sl neutrino} mass are smaller than the
tree-level contribution. In this case it is possible to give a simple
approximate formula for the composition of the third neutrino mass
eigenstate,
\be{AppMix}
U_{\alpha,3} = Sin(Atan(\frac{\Lambda_{\alpha}}
                        {\sqrt{\sum_{\beta \ne \alpha}\Lambda_{\beta}^2}}))
\ee

%
%

\vskip-55mm
\epsfysize=100mm
\epsfxsize=110mm
\epsfbox{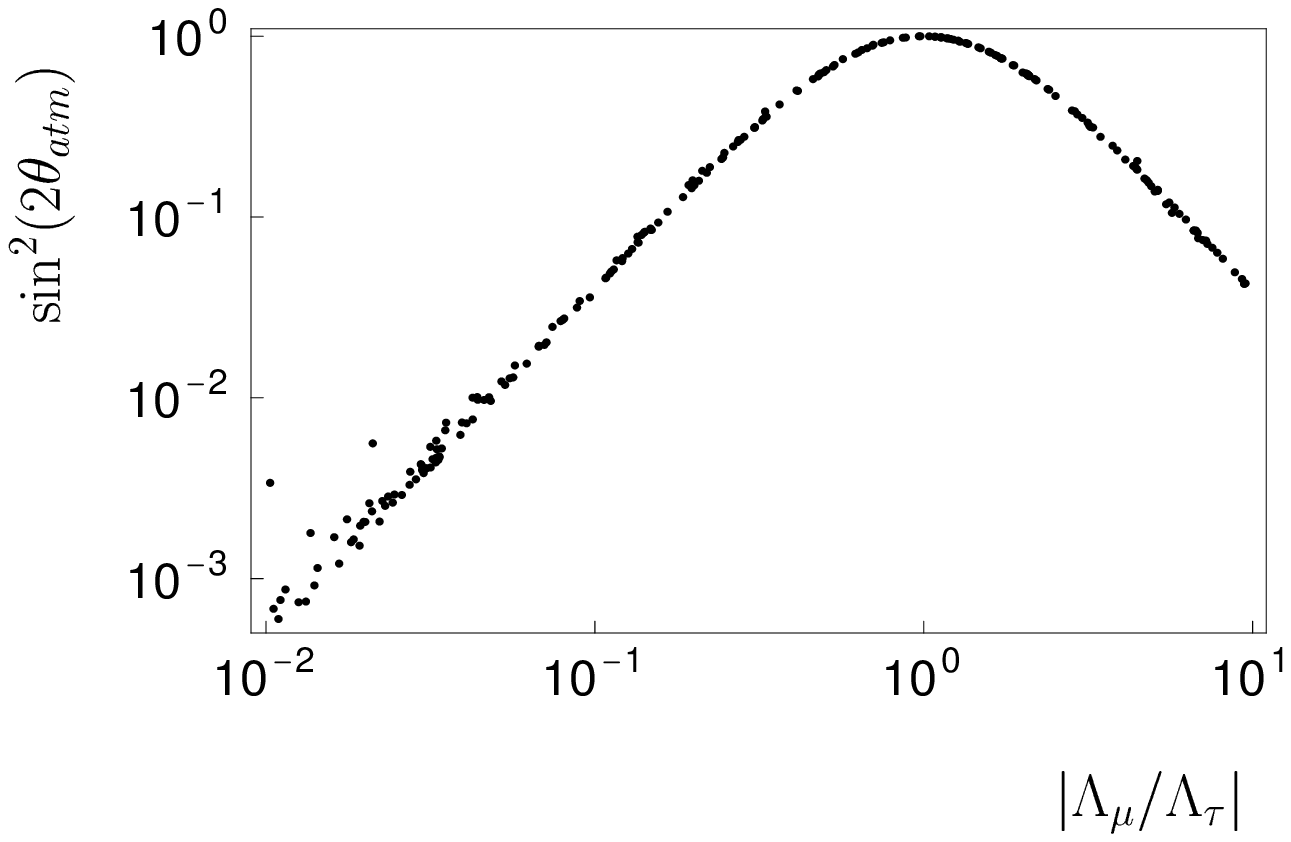}
%
\vskip20mm
\noindent
{\bf Figure 1: }{\it The atmospheric angle as function of
$|\Lambda_{\mu}/\Lambda_{\tau}|$, for $|\epsilon_i| = \epsilon$ and
$\Lambda_{e} = 0.1 \Lambda_{\tau}$. Here $\epsilon^2/\Lambda \lsim
0.1$, since larger values lead to larger scatter for very small
$|\Lambda_{\mu}/\Lambda_{\tau}|$. Maximality of atmospheric mixing is
only possible for $|\Lambda_{\mu}| \simeq |\Lambda_{\tau}|$.}
\bigskip
%
%

%
Accounting for the atmospheric neutrino anomaly requires that the
$\nu_{\mu}-\nu_{\tau}$ mixing be large, with little effect of
$\nu_{e}$ in the atmospheric neutrino oscillations. Fitting for the
atmospheric neutrino data then fixes $|\Lambda_{\mu}/\Lambda_{\tau}|$
through this simple equation. These parameters are dynamically
determined since they involve Higgs and sneutrino VEVS obtained from
the scalar potential.  In fact, the $|\Lambda_{\alpha}|$ parameters
are simply proportional to the sneutrino vacuum expectation values in
the basis where the bilinear term in the superpotential is ``rotated
away'' in favour of a tri-linear one~\cite{epsrad}. As an illustration
Figure 1 shows the $\nu_{\mu}-\nu_{\tau}$ angle as a function of
$|\Lambda_{\mu}/\Lambda_{\tau}|$ for $\Lambda_e \simeq 0.1
\Lambda_{\mu}$ for an otherwise random variation of parameters.
Clearly the condition $|\Lambda_{\mu}|=|\Lambda_{\tau}|$ is sufficient
to ensure near maximal mixing, as long as $\Lambda_e$ is somewhat
suppressed.

One immediate consequence of the smallness of loop with respect to
tree contributions is that the absolute \rp scale is then fixed by the
atmospheric neutrino mass scale. For the above choice of sampling one
has $|\Lambda| \simeq 0.03 - 0.25$ GeV$^2$.  While this value is
surely smaller than the weak scale, it may arise naturally in models
where the sneutrino VEVS are generated radiatively~\cite{epsrad}.

With the magnitude of \rp parameters fixed by the atmospheric neutrino
problem, the question arises, whether the loop-induced oscillation
parameters, mass splitting and angle, are in the right range for
either the vacuum or the MSW solution to the solar neutrino
problem. Since the ratio of the loop masses to the tree-level mass
depends on the relative size of the bi-linear \rp parameters with
respect to the alignment vector $\Lambda$ this can not in general be
predicted in the bi-linear \rp model. We have found, however, that
with our assumption of generation-independent bi-linear parameters
$\epsilon_i$ there should be a relative sign between the dynamically
determined \rp $\Lambda$ parameters, i.e. $\Lambda_{\mu} \simeq -
\Lambda_{\tau}$.  
Figure 2 shows how, having fixed the 
$\Lambda_i$ by the atmospheric neutrino problem, the solar angle is
determined under the above sign assumption as a function of
$\epsilon_{e}/\epsilon_{\mu}$.

%
%

\vskip-55mm
\epsfysize=100mm
\epsfxsize=110mm
\epsfbox{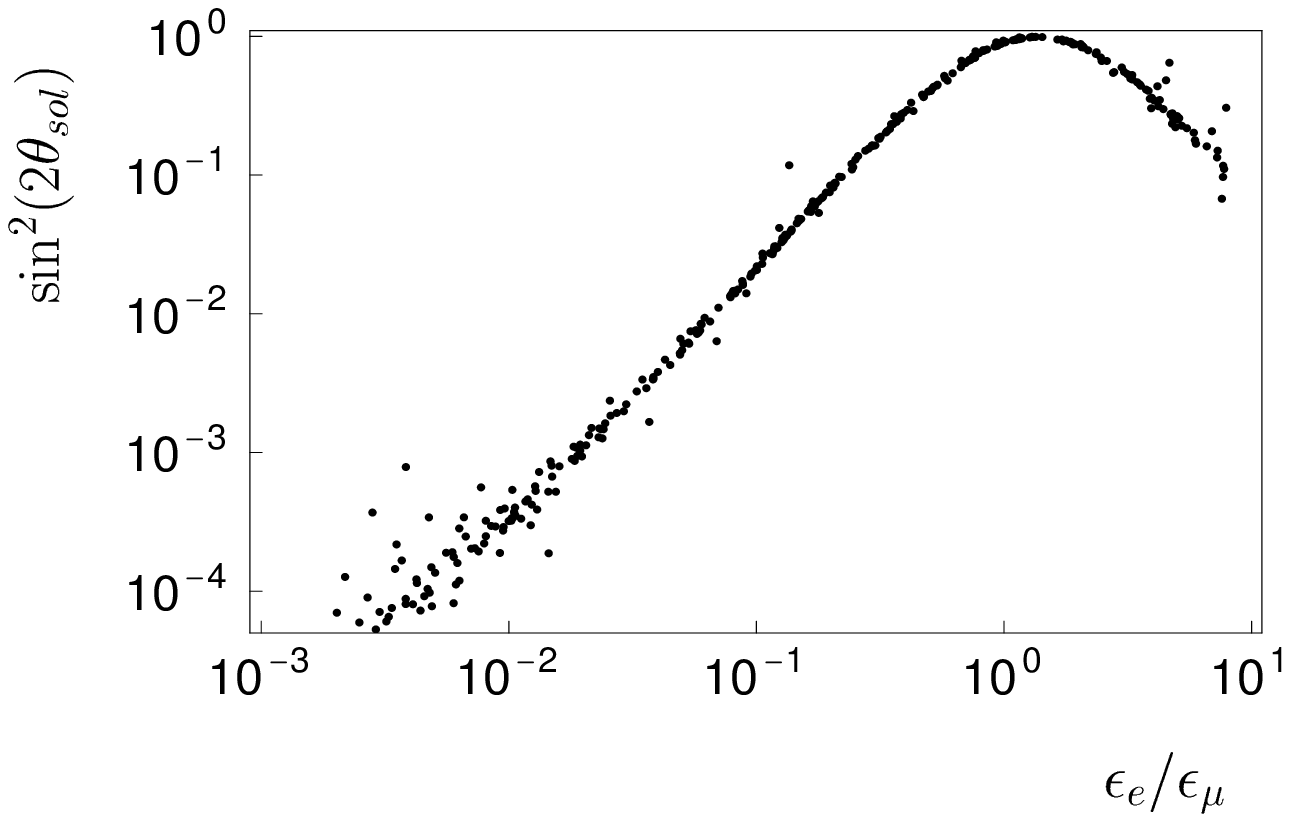}

\vskip15mm
\noindent
{\bf Figure 2: }{\it The solar angle as function of 
$\epsilon_e/\epsilon_{\mu}$, for $\epsilon_{\mu}=\epsilon_{\tau}$  
and $\Lambda_{\mu}=\Lambda_{\tau}$, but $\Lambda_{e}=0.1 \Lambda_{\mu}$, 
applying the condition: 
$(\Lambda_{\mu}/\Lambda_{\tau})\times (\epsilon_{\mu}/\epsilon_{\tau}) 
\le 0$. Maximality of solar mixing is only possible
for $\epsilon_{\mu} \simeq \epsilon_{e}$.}
\bigskip
%
%

As for the solar neutrino scale we show
in fig. 3 $\Delta m^2_{12}$ versus $\epsilon^2/|\Lambda |$, where
$\epsilon^2 = \sum_i \epsilon_i^2$. As is seen from the figure, for
fixed tree-level mass the loop masses depend strongly on this
quantity. Large values give $\Delta m^2_{12}$ in the MSW range, while
low values could give vacuum solutions to the solar neutrino
problem.

While a certain amount of ``alignment'' is needed, for masses
in the MSW range, the model by itself does not prefer one solution
over the other. It is also clear that bi-maximal neutrino mixing is
generated in the bilinear \rp MSSM - independent of the actual values
of SUSY parameters, if (i) the \rp bi-linear terms are (nearly)
generation blind, (ii) $\tan\beta \lsim 10$ implying that the lightest
CP even Higgs boson mass lies below 115 GeV or so, and (iii)
$\Lambda_{\mu} \simeq - \Lambda_{\tau}$, as long as $\Lambda_{e}$ is
somewhat smaller than the $\Lambda_{\mu}$ and $\Lambda_{\tau}$.  We
have checked explicitly that (iii) arises dynamically by minimizing
the scalar potential of the theory.

%
%
\vskip-55mm
\epsfysize=100mm
\epsfxsize=110mm
\epsfbox{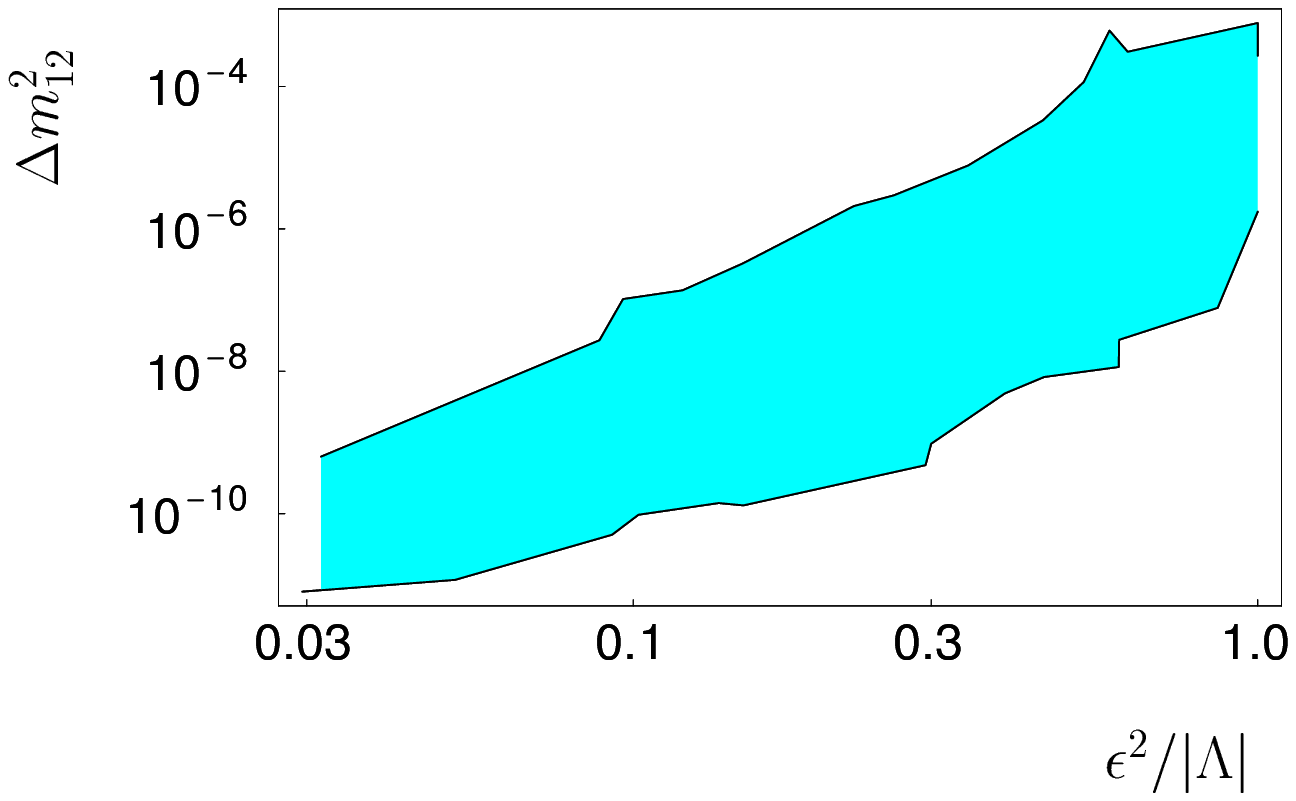}

\vskip15mm
\noindent
{\bf Figure 3: }{\it $\Delta m^2_{12}$ as function of 
$\epsilon^2/|\Lambda|$. Low values lead to neutrino masses 
in the just-so range, whereas high values give $\Delta m^2_{12}$ 
in the range of the MSW solution.}
\bigskip
%
%


Finally, apart from the possible detection of lightest CP even Higgs
boson (mass below 115 GeV or so), we would like to point out that,
despite the smallness of neutrino masses R-parity violation is
observable at accelerators through the observation of the decay of the
Lightest Supersymmetric Particle (LSP), typically a neutralino. For
example for a LSP mass of about 50 to 60 GeV the decay will occur
inside typical LEP, Tevatron and LHC detectors (the neutralino decay
length can be of the order of one meter or so), diluting the missing
momentum signal to a maximum of 15\% of the MSSM expectations. The LSP
decay will give rise to high multiplicity events, providing an
unambiguous test of the model. This is specially so if branching
ratios are measured. In order to demonstrate this correlation we have
calculated the ratio of semi-leptonic branching ratios of the LSP into
muons and taus. We found the striking result that the correlation
depicted in fig 1 which is required by the atmospheric neutrino
anomaly is mapped into a well-defined correlation for the ratio of
semi-leptonic LSP branching ratios into muons and taus (Figure 4).
Note that this correlation holds for the semi-leptonic decays, despite
the fact that there are many scalar boson exchanges contributing to
the LSP decay! The case where the LSP is heavier than the W and has
2--body decays mainly to W and Z was considered in
ref.~\cite{Roy:1997bu} who found a similar correlation in the
approximation where the LSP decay into lightest neutral supersymmetric
Higgs boson was neglected. The result we show in Figure 4 is general,
independently of the neutralino mass, and provides a powerful way to
probe the solution of the atmospheric neutrino anomaly and opening the
potential to measure the related neutrino angles at high energy
accelerators!

%
%
\vskip-25mm
\epsfysize=70mm
\epsfxsize=90mm
\epsfbox{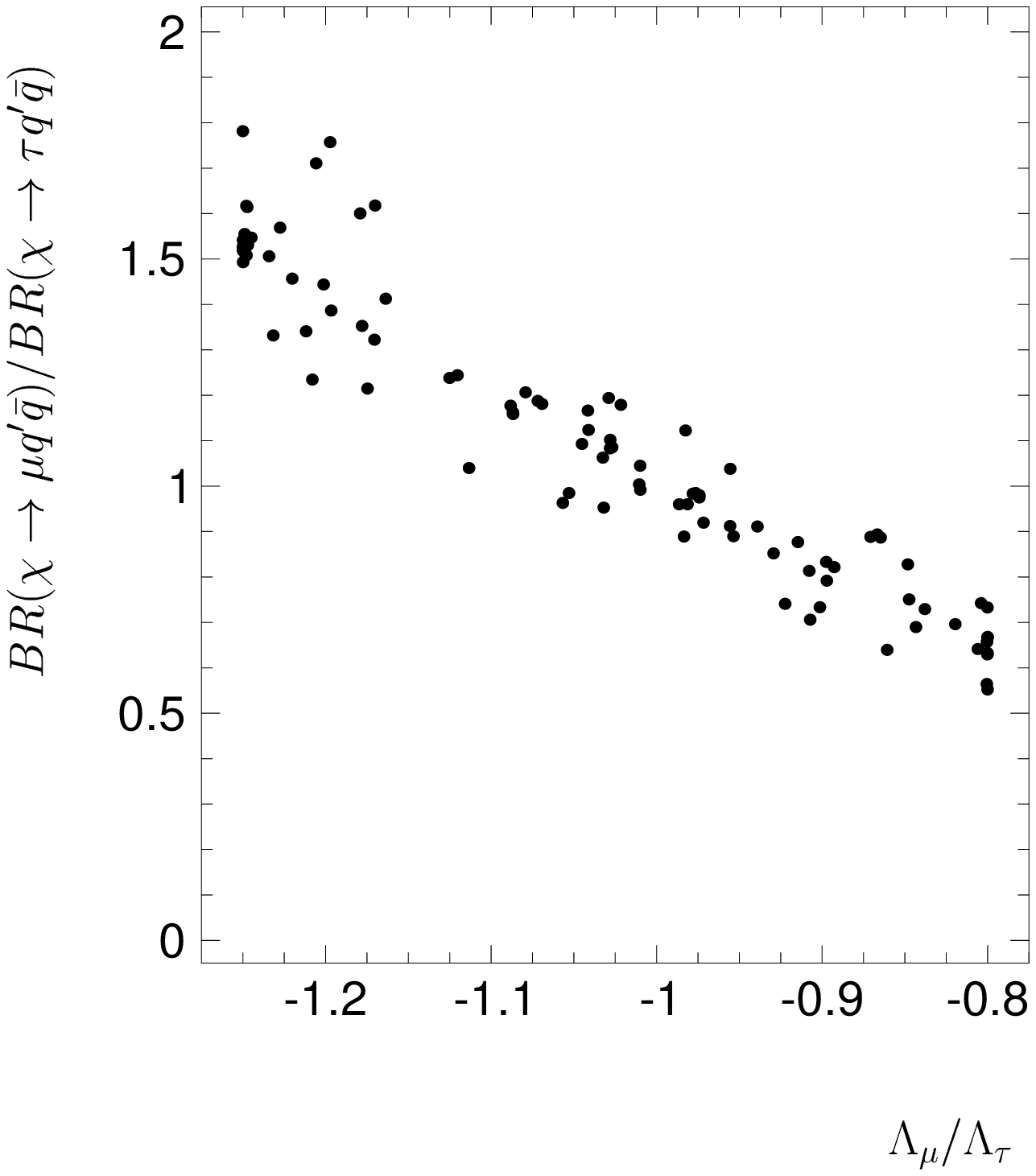}

\vskip25mm
\noindent
{\bf Figure 4: }{\it Ratio of branching ratios for semileptonic LSP
decays into muons and taus: $BR(\chi \to \mu q' \bar q)/ BR(\chi \to
\tau q' \bar q$) as function of $\Lambda_\mu/\Lambda_\tau$. }

\bigskip
%
%

\vskip .3cm

This work was supported by DGICYT under grant PB95-1077, by Accion
Integrada Hispano Austriaca HU97-46 and by the TMR network grant
ERBFMRXCT960090 of the European Union. W. P. and M. H. were supported
respectively by the Spanish MEC (contract N. SB97-BU0475382) and by a 
Marie Curie TMR fellowship of
the European Union, contract N. ERBFMBICT983000. M.A.D. was partly
supported by the D.O.E. under DE-FG02-97ER41022

\end{document}